\documentclass[prl,aps,twocolumn,showpacs]{revtex4}

\usepackage{pxfonts}
\usepackage{graphicx}
\begin{document}

\title{Indirect excitons in van der Waals heterostructures at room temperature}

\author{E.V. Calman, M.M. Fogler, L.V. Butov}
\affiliation{Department of Physics, University of California at San Diego, La Jolla, CA 92093-0319, USA}
\author{S. Hu, A. Mishchenko, A.K. Geim}
\affiliation{School of Physics and Astronomy, University of Manchester, Manchester M13 9PL, UK}

\date{\today}

\begin{abstract}
Indirect excitons (IXs) in van der Waals transition-metal dichalcogenide (TMD) heterostructures are characterized by a high binding energy making them stable at room temperature and giving the opportunity for exploring fundamental phenomena in excitonic systems and developing excitonic devices operational at high temperatures. We present the observation of IXs at room temperature in van der Waals TMD heterostructures based on monolayers of MoS$_2$ separated by atomically thin hexagonal boron nitride. The IXs realized in the TMD heterostructure have lifetimes orders of magnitude longer than lifetimes of direct excitons in single-layer TMD, and their energy is gate controlled.
\end{abstract}


\date{\today}

\maketitle

An indirect exciton (IX) is composed of an electron and a hole confined in spatially separated quantum well layers. IXs can be realized in coupled quantum well (CQW) semiconductor heterostructures. Long lifetimes of IXs allow them to cool below the temperature of quantum degeneracy~\cite{Lozovik1976, Butov2001}. Experimental studies of degenerate Bose gases of IXs were performed so far in GaAs CQW structures where quantum degeneracy was achieved in the temperature range of few Kelvin. The findings include spontaneous coherence and condensation of excitons~\cite{High2012}, long-range spin currents and spin textures~\cite{High2013}, spatially modulated exciton state~\cite{Yang2015}, and perfect Coulomb drag~\cite{Nandi2012}. Furthermore, IX energy, lifetime, and flux can be controlled by voltage that is explored for the development of excitonic devices. Excitonic devices with IXs were demonstrated so far at temperatures below $\sim 100$~K. These devices include traps, lattices, conveyers, and ramps, which are used for studying basic properties of cold IXs, as well as excitonic transistors, routers, and photon storage devices, which hold the potential for creating excitonic signal processing devices and excitonic circuits, a review of excitonic devices can be found in~\cite{Butov2017}.

A finite exciton binding energy $E_{\rm ex}$ limits the operation temperature of excitonic devices. Excitons exist in the temperature range roughly below $E_{\rm ex} / k_{\rm B}$ ($k_{\rm B}$ is the Boltzmann constant)~\cite{Chemla1984}. Furthermore, the temperature of quantum degeneracy, which can be achieved with increasing density before excitons dissociation to electron-hole plasma, also scales proportionally to $E_{\rm ex}$~\cite{Fogler2014}. These considerations instigate the search for material systems where IXs have a high binding energy and, as a result, can provide the medium for the realization of high-temperature coherent phenomena and excitonic devices. 

IXs were explored in various III-V and II-VI semiconductor QW heterostructures based on GaAs~\cite{Butov2001, High2012, High2013, Yang2015, Nandi2012, Butov2017, Chemla1984, Islam1987, Alexandrou1990}, AlAs~\cite{Zrenner1992, Grosso2009}, InGaAs~\cite{Butov1995}, GaN~\cite{Lefebvre2004, Fedichkin2015, Fedichkin2016}, and ZnO~\cite{Morhain2005, Kuznetsova2015}. Among these materials, IXs are more robust in the ZnO structures where their binding energy is about 30~meV~\cite{Morhain2005}. Proof of principle for the operation of IX switching devices was demonstrated at temperatures up to $\sim 100$~K in AlAs/GaAs CQW~\cite{Grosso2009} where the IX binding energy is about $\sim 10$~meV~\cite{Zrenner1992}. Studies of IXs in III-V and II-VI semiconductor materials continue to attract intense interest. 

Van der Waals structures composed of atomically thin layers of transition metal dichalcogenides (TMD) offer an opportunity to realize artificial materials with designable properties, forming a new platform for studying basic phenomena and developing optoelectronic devices~\cite{Geim2013}. TMD heterostructures allow IXs with remarkably high binding energies~\cite{Fogler2014, Ye2014, Chernikov2014}, much higher than in III-V or II-VI semiconductor heterostructures. Therefore, IXs in TMD heterostructures open the opportunity to realize room-temperature excitonic devices and explore high-temperature quantum degenerate Bose gases of IXs. 

The experimental approaches to the realization of IXs in TMD materials involve two kinds of heterostructures. In type I TMD heterostructures with direct gap alignment, the electron and hole layers are spatially separated by a barrier layer. Such type I structures with MoS$_2$ forming the QW layers and a hexagonal boron nitride (hBN) forming the barrier were considered in~\cite{Fogler2014, Calman2016}. These heterostructures are similar to GaAs/AlGaAs CQW heterostructures where GaAs forms QW layers and AlGaAs forms the barrier~\cite{Butov2001, High2012, High2013, Yang2015, Nandi2012, Butov2017, Islam1987, Alexandrou1990}. In type II TMD heterostructures with staggered band alignment, the electron and hole layers form in different adjacent TMD materials such as single-layer MoSe$_2$ and WSe$_2$~\cite{Rivera2015, Rivera2016, Nagler2017}, MoS$_2$ and WSe$_2$~\cite{Fang2014, Latini2017}, MoS$_2$ and WS$_2$~\cite{Hong2014, Palummo2015}, MoSe$_2$ and WS$_2$~\cite{Bellus2015}, and MoS$_2$ and MoSe$_2$~\cite{Mouri2017}. These heterostructures are similar to AlAs/GaAs CQW heterostructures where electrons and holes are confined in adjacent AlAs and GaAs layers, respectively~\cite{Zrenner1992, Grosso2009}. 

Here, we report on the realization of IXs in type I TMD heterostructures at room temperature. This was achieved using the previously demonstrated approach with MoS$_2$/hBN type-I CQW~\cite{Calman2016} combined with an improved structure design and detected using time-resolved optical spectroscopy.

\begin{figure}[b]
\includegraphics[width=8.5cm]{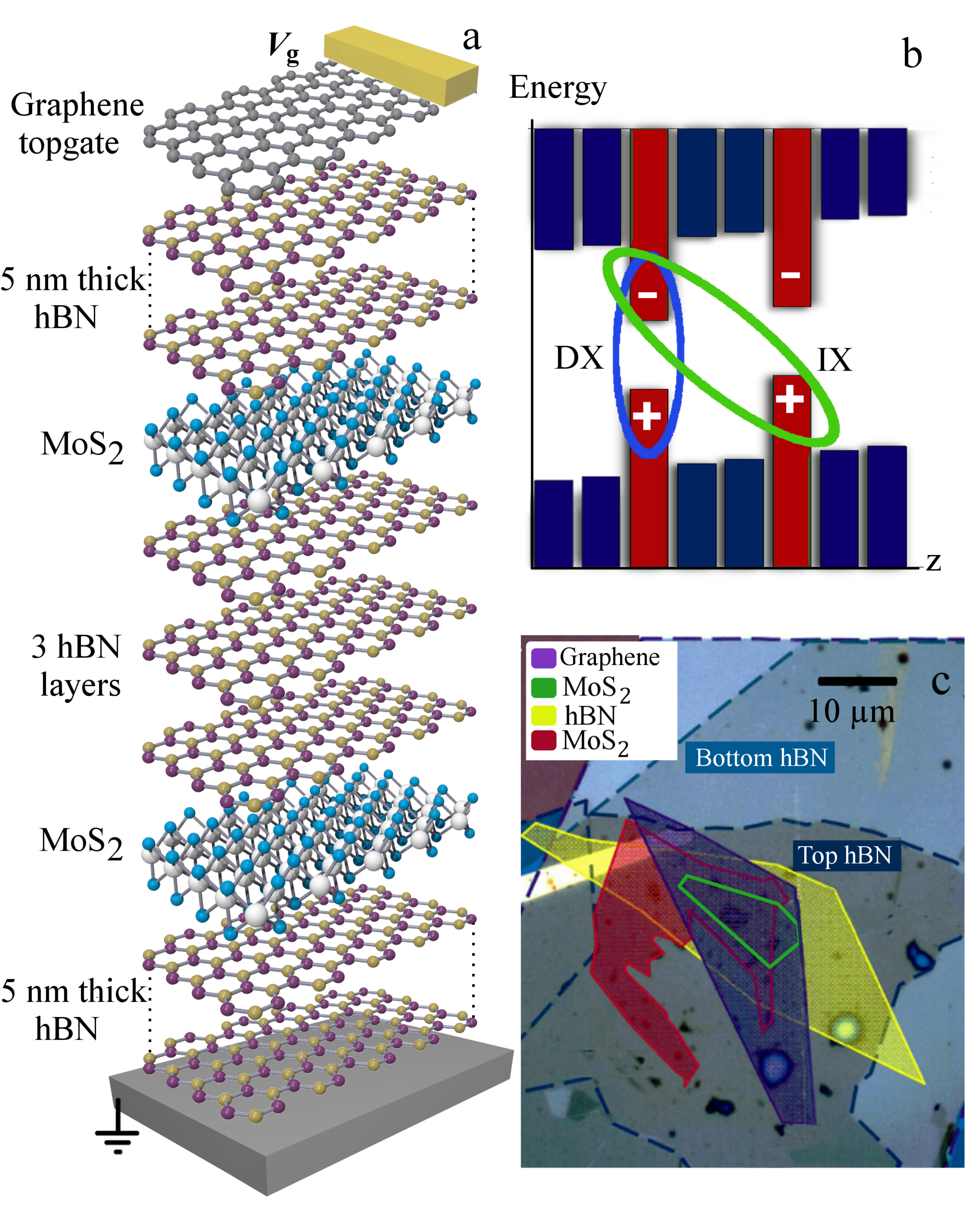}
\caption{The coupled quantum well van der Waals heterostructure layer (a) and energy-band (b) diagrams. The ovals indicate a direct exciton (DX) and an indirect exciton (IX) composed of an electron ($-$) and a hole ($+$). (c) Microscope image showing the layer pattern of the device. 
} \label{fig:spectra}
\end{figure}

The structure studied here was assembled by stacking mechanically exfoliated two-dimensional crystals on a graphite substrate, which acts as a global backgate (Fig.~1a). The top view of the device showing the contours of different layers is presented in Fig.~1c. The CQW is formed where the two MoS$_2$ monolayers, separated by three hBN layers, overlap. IXs are formed from electrons and holes in different MoS$_2$ layers (Fig.~1b). The top and bottom 5~nm thick hBN serve as dielectric cladding layers. Voltage $V_{\rm g}$ applied between the top graphene layer and the backgate is used to create the bias across the CQW structure. The thickness of hBN cladding layers is much smaller than in our previous CQW TMD device~\cite{Calman2016}. This allowed us to achieve a much higher electric field across the structure for the applied voltage and, in turn, realize effective control of IX energy by voltage as described below.  

\begin{figure}[b]
\includegraphics[width=8.5cm]{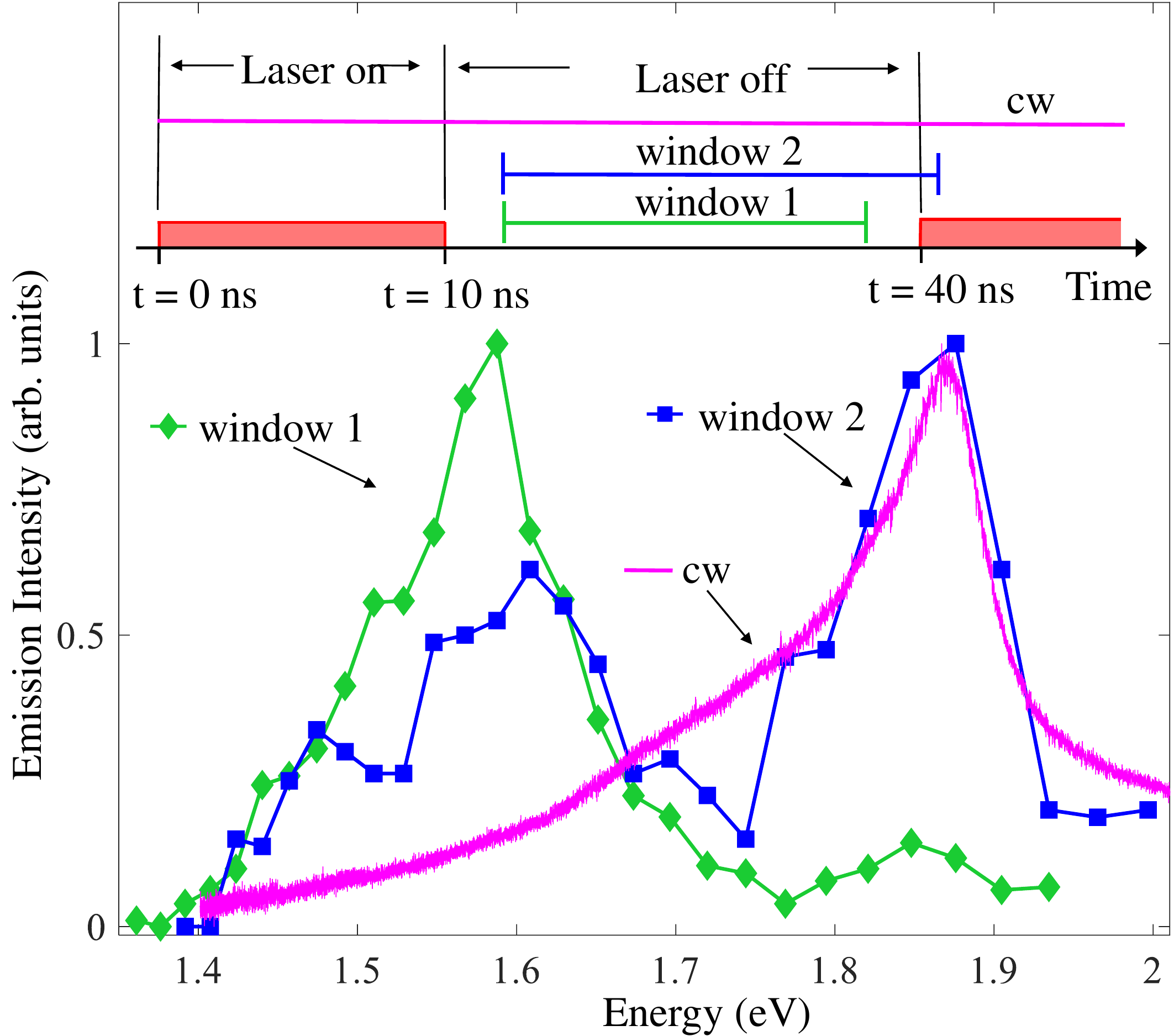}
\caption{Spectrum taken in a time-integration window after the laser pulse (window 1) when most of short-lifetime DXs recombine (green diamonds). The spectrum shows emission of long-lifetime IXs. The laser profile and signal integration windows are shown above. The laser has a pulse duration of 10~ns and a period of 40~ns. Window 2 presents emission from a combination of the laser off and the laser on in a ratio that shows both IX and DX (blue squares). cw spectrum (magenta line) is dominated by direct recombination. $T = 300$~K. $V_{\rm g} = 0$.
} \label{fig:spectra}
\end{figure}

The excitons were generated by a semiconductor laser with excitation energy $E_{\rm ex} = 3.07$~eV. In continuous wave (cw) experiments, photoluminescence (PL) spectra were measured using a spectrometer with resolution 0.2~meV and a liquid-nitrogen-cooled CCD. In all time-resolved experiments, the laser pulses had a rectangular shape with the duration 10--50~ns, period 40--200~ns, and edge sharpness $\sim 0.5$~ns (Fig.~2). The laser was focused to a $\sim 2$~$\mu$m spot. In time-resolved PL kinetics and spectrum measurements, the emitted light was filtered by an interference filter or diffracted by the spectrometer, respectively, and then detected by a photomultiplier tube and time correlated photon counting system. In time-resolved imaging experiments, the emitted light was filtered by an interference filter and detected by a liquid-nitrogen-cooled CCD coupled to a PicoStar HR TauTec time-gated intensifier. The measurements were performed in a $^4$He atmosphere at room temperature ($T \approx 300$~K) and in a liquid $^4$He cryostat at 2~K.

IXs dominate the emission spectrum measured after the laser excitation pulse (Fig.~2). At the time delays exceeding the DX recombination times, most of DXs have recombined, and so the recombination of IXs, which have a much longer lifetime, is not masked by the DX recombination. Both short-lifetime DX and long-lifetime IX emission lines are observed in the spectrum measured in the time-window between the laser pulses and the first $\approx 2$~ns of the laser pulse (Fig.~2). As the fraction of time corresponding to the laser pulse grows, the relative intensity of the DX emission increases. In the cw regime, where the laser is permanently on, DXs dominate the spectrum due to their higher oscillator strength (Fig.~2). 

\begin{figure}
\begin{center}
\includegraphics[width=7.5cm]{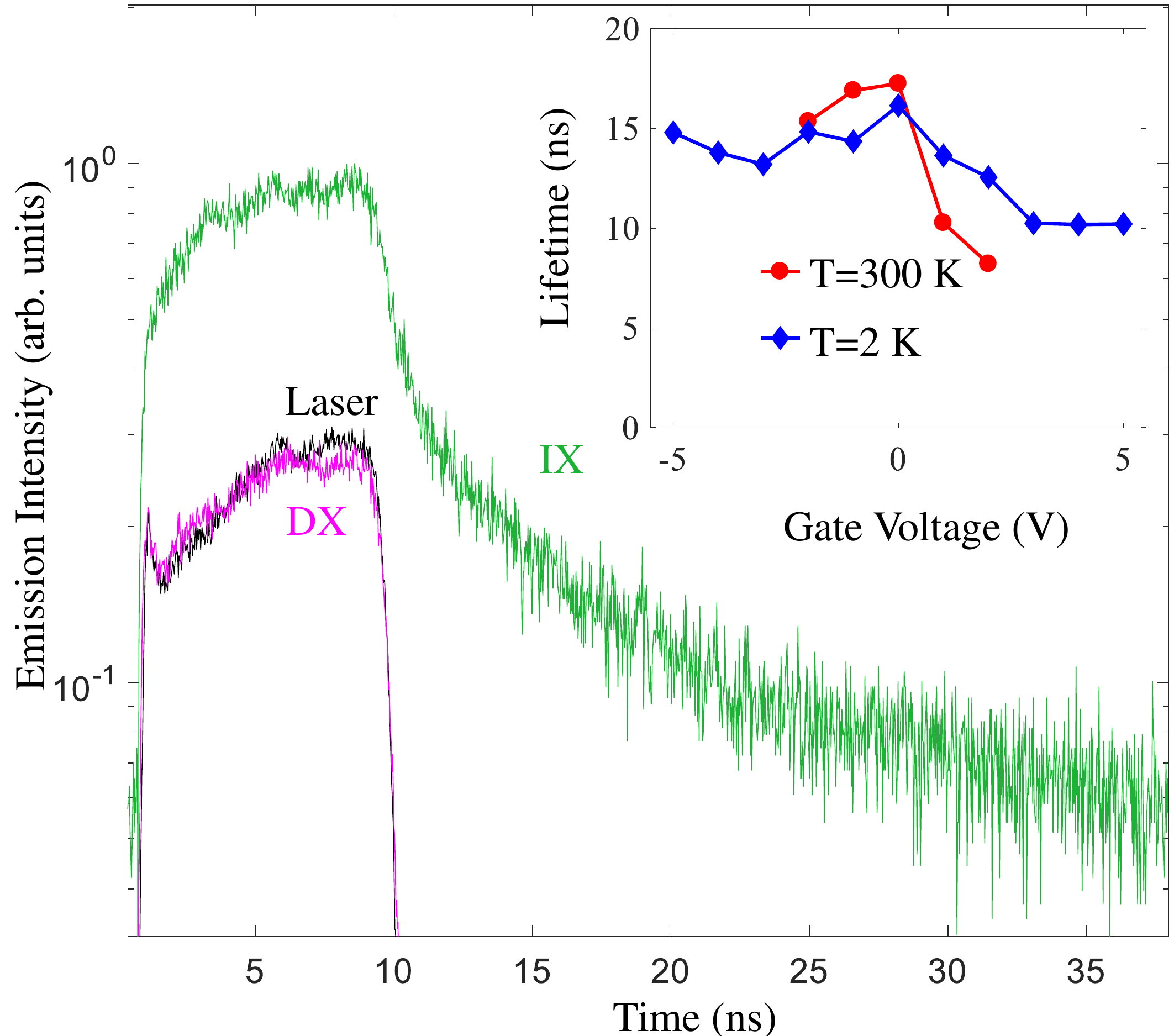}
\caption{\label{fig3} {Emission kinetics at energies of 1.46--1.65~eV corresponding to the IX spectral range (green), 1.89~eV corresponding to the DX spectral range (magenta), and 3.07~eV corresponding to the excitation laser (black) at $T = 300$~K and $V_{\rm g} = 0$. IX lifetimes are shown in the inset as a function of gate voltage $V_{\rm g}$ for $T = 2$~and 300~K, see text. The laser excitation has a pulse duration of 10~ns and a period of 40~ns.}}
\end{center}
\end{figure}

The IX emission kinetics is presented in Fig.~3. The time resolution of the experimental system including the pulse generator, the laser, the photomultiplier, and the time correlated photon counting system is approximately 0.5~ns as seen from the decay kinetics measured at $E_{\rm ex} = 3.07$~eV. The DX decay kinetics measured at the DX line peak $E_{\rm DX} = 1.89$ eV closely follows the excitation laser decay indicating that the DX lifetime is shorter than the 0.5~ns experimental resolution. The decay kinetics in the IX spectral range $1.46 - 1.65$~eV shows a double-exponential decay (Fig.~3). Its faster component is determined by the decay of low-energy DX states, which appear in the IX spectral range due to the spectral broadening of the DX line (Fig.~2). The slower component is determined by the IX decay (Fig.~3). The IX lifetime $\sim 10$~ns is orders of magnitude longer than the DX lifetime (Fig.~3) and is controlled by gate voltage $V_{\rm g}$ over a range of several ns (Fig.~3 insert). The voltage dependence of the IX lifetime has two characteristic features. First, it reduces at positive $V_{\rm g}$ where the IX energy approaches the DX energy (Fig.~5). Second, it has a local maximum around $V_{\rm g} = 0$. Both these features were also observed for IXs in GaAs heterostructures~\cite{Grosso2009}. The former can be attributed to the increase of the overlap of electron and hole wave functions with approaching the direct regime. The latter may result from the suppression of the leakage currents through the CQW layers at zero bias. The realization of the indirect regime, where the IXs are lower in energy than DXs, already at $V_{\rm g} = 0$ indicates an asymmetry of the device, presumably due to unintenional doping.

Regarding the physical mechanism that governs the IX lifetime in the studied heterostructure we can say the following. In general, this lifetime is limited by tunneling through the hBN spacer. However, direct tunneling across the entire thickness $3 \times 0.33 = 1$~nm of the spacer should be prohibitively slow. The tunneling action and tunneling probability can be estimated to be $S \sim 12$ and $\exp(-2S) \sim 10^{-11}$, respectively, assuming the potential barrier of height 2~eV and the carrier mass $m_b \sim 0.5$ inside the barrier (similar to \cite{Fogler2014}). Therefore, we surmise that the IX recombination involves transmission through some midgap defects in the spacer~\cite{Britnell2012}.

\begin{figure}
\begin{center}
\includegraphics[width=7.5cm]{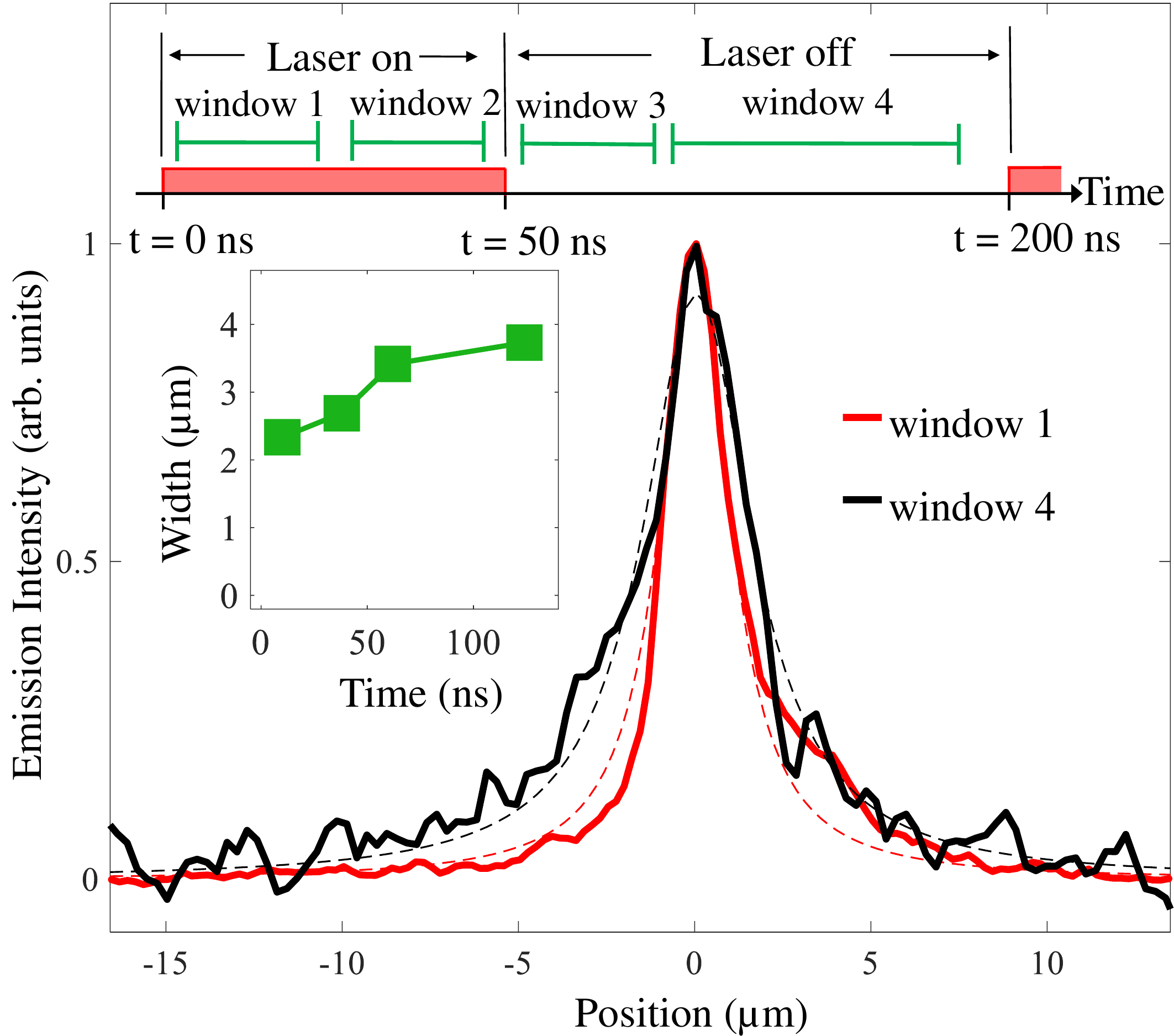}
\caption{\label{fig4} {Spatial width of the emission at the IX energy 1.46--1.65 eV measured in time windows shown above. The time evolution of the exciton emission width extracted by fitting to lorentzian profiles (dashed lines) is shown in the inset. The excitation laser has a pulse duration of 50~ ns and a period of 200~ns. $T = 300$~K. $V_{\rm g} = 0$}.}
\end{center}
\end{figure}

Figure ~4 shows the spatial profiles of IX emission. The width of the emission profiles determined by a fit to Lorentzian distribution (dashed lines in Fig.~4) is shown in insert to Fig.~4 as a function of time. The emission profiles after the laser excitation pulse have wider spatial distributions than during the pulse. During the laser excitation pulse, the low-energy tail of short-lifetime DX emission strongly contributes to the emission in the IX spectral region (Fig.~2). After the pulse, DXs decay quickly and the emission is dominated by the long-lifetime IXs. Diffusion of IXs away from the laser excitation spot during their long lifetime contributes to the wider spatial profiles of IX emission. The increase in emission width after the pulse end $l$ can be used for estimating an upper bound on the IX diffusion coefficient $D$. For $l \sim1$~$\mu$m (Fig.~4) and IX lifetime $\tau \sim 10$~ns (Fig.~3) this gives $D \sim l^2 / \tau \sim 1$~cm$^2$/s.  

\begin{figure}
\begin{center}
\includegraphics[width=8cm]{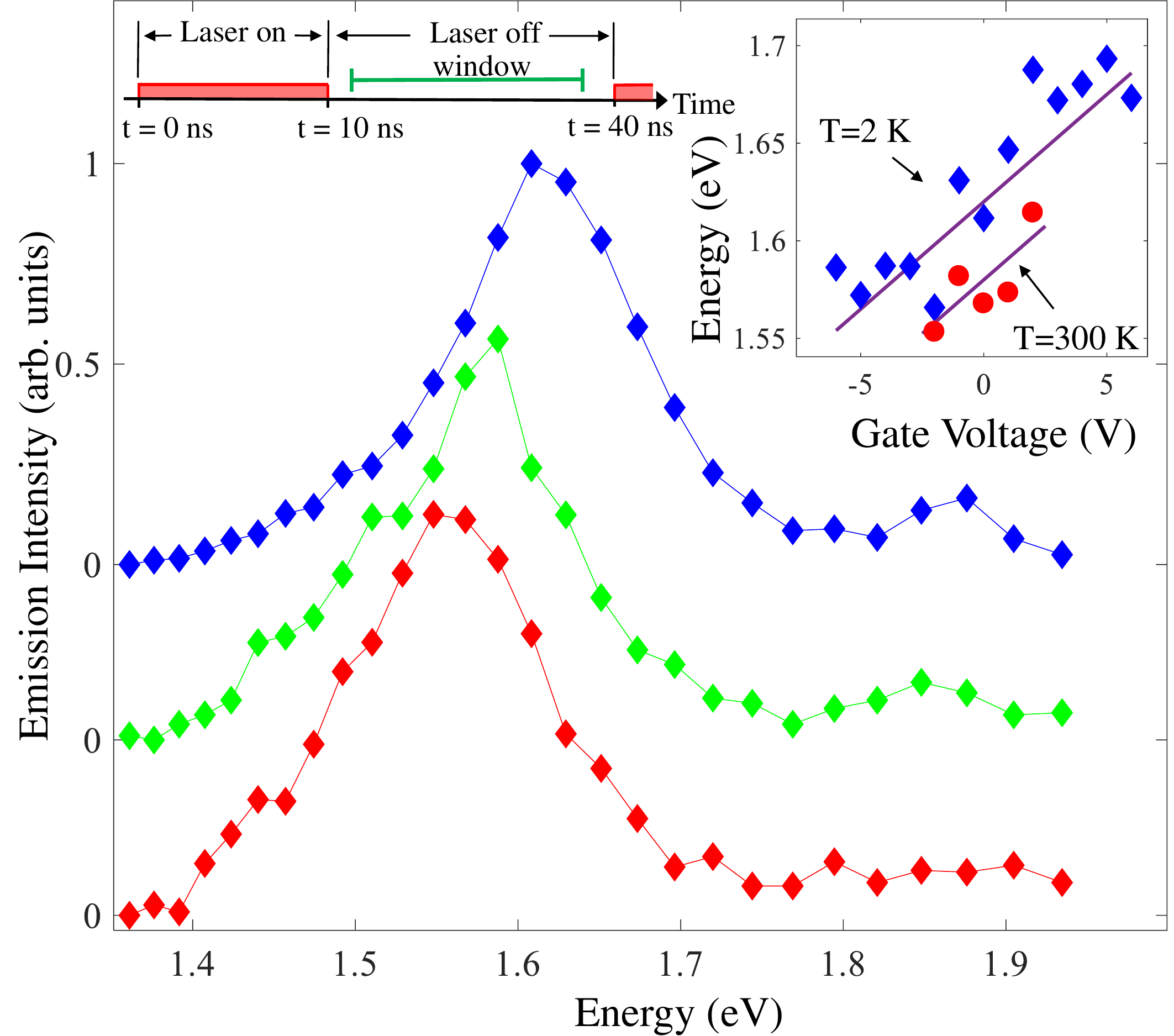}
\caption{\label{fig5} {Spectra at a delay after the laser excitation pulse in the time window shown above at gate voltages $V_{\rm g} = - 2, 0,$~and 2~V at $T = 300$~K. Gate voltage dependence of IX energies at $T = 300$~and 2~K is shown in the insert. The excitation  laser has a pulse duration of 10~ns and a period of 40~ns.}}
\end{center}
\end{figure}

Figure ~5 shows control of the IX energy by gate voltage. The energy of the long-lifetime emission line shifts by about 120~meV at cryogenic temperatures and by about 60~meV at room temperature. No leakage current or sample damage was detected at cryogenic temperatures in the measurements at applied voltages up to $\pm 6$~V, which is typical for thin hBN that can withstand electric fields of about 0.5 V/nm~\cite{Britnell2012}. However, at room temperature, applying $\pm 3$~V led to the appearance of leakage current through the device and the reduced device resistivity persisted after lowering $V_{\rm g}$. This limited the maximum applied voltage and, in turn, the IX energy shift at room temperature.

In summary, IXs were observed at room temperature in van der Waals MoS$_2$/hBN heterostructure. The IXs have long lifetimes, orders of magnitude longer than lifetimes of direct excitons in single-layer MoS$_2$, and their energy is controlled by voltage at room temperature.

These studies were supported by DOE Office of Basic Energy Sciences under award DE-FG02-07ER46449 and NSF Grant No. 1640173 and NERC, a subsidiary of SRC, through the SRC-NRI Center for Excitonic Devices.


\begin{references}

\bibitem{Lozovik1976}
Yu.E. Lozovik, V.I. Yudson, A new mechanism for superconductivity: pairing between spatially separated electrons and holes. {\it Sov. Phys. JETP} {\bf 44}, 389-397 (1976).

\bibitem{Butov2001}
L.V. Butov, A.L. Ivanov, A. Imamoglu, P.B. Littlewood, A.A. Shashkin, V.T. Dolgopolov, K.L. Campman, A.C. Gossard, Stimulated Scattering of Indirect Excitons in Coupled Quantum Wells: Signature of a Degenerate Bose-Gas of Excitons, {\it Phys. Rev. Lett.} {\bf 86}, 5608-5611 (2001).

\bibitem{High2012}
A.A. High, J.R. Leonard, A.T. Hammack, M.M. Fogler, L.V. Butov, A.V. Kavokin, K.L. Campman, A.C. Gossard, Spontaneous coherence in a cold exciton gas, {\it Nature} {\bf 483}, 584-588 (2012).

\bibitem{High2013}
A.A. High, A.T. Hammack, J.R. Leonard, Sen Yang, L.V. Butov, T. Ostatnick{\'y}, M. Vladimirova, A.V. Kavokin, T.C.H. Liew, K.L. Campman, A.C. Gossard, Spin currents in a coherent exciton gas, {\it Phys. Rev. Lett.} {\bf 110}, 246403 (2013).

\bibitem{Yang2015}
Sen Yang, L.V. Butov, B.D. Simons, K.L. Campman, A.C. Gossard, Fluctuation and commensurability effect of exciton density wave, {\it Phys. Rev. B} {\bf 91}, 245302 (2015).

\bibitem{Nandi2012}
D. Nandi, A.D.K. Finck, J.P. Eisenstein, L.N. Pfeiffer, K.W. West, Exciton condensation and perfect Coulomb drag, {\it Nature} {\bf 488}, 481-484 (2012).

\bibitem{Butov2017}
L.V. Butov, Excitonic devices, {\it Superlattices Microstruct.} {\bf 108}, 2-26 (2017).

\bibitem{Chemla1984}
D.S. Chemla, D.A.B. Miller, P.W. Smith, A.C. Gossard, W. Wiegmann, Room temperature excitonic nonlinear absorption and refraction in GaAs/AlGaAs multiple quantum well structures, {\it IEEE J. Quantum Electron.} {\bf 20}, 265-275 (1984).

\bibitem{Fogler2014}
M.M. Fogler, L.V. Butov, K.S. Novoselov, High-temperature superfluidity with indirect excitons in van der Waals heterostructures, {\it Nat. Commun.} {\bf 5}, 4555 (2014).

\bibitem{Islam1987}
M.N. Islam, R.L. Hillman, D.A.B. Miller, D.S. Chemla, A.C. Gossard, J.H. English, Electroabsorption in GaAs/AlGaAs coupled quantum well waveguides, {\it Appl. Phys. Lett.} {\bf 50}, 1098-1100 (1987).

\bibitem{Alexandrou1990}
A. Alexandrou, J.A. Kash, E.E. Mendez, M. Zachau, J.M. Hong, T. Fukuzawa, Y. Hase, Electric-field effects on exciton lifetimes in symmetric coupled GaAs/Al$_{0.3}$Ga$_{0.7}$As double quantum wells, {\it Phys. Rev. B} {\bf 42}, 9225-9228 (1990).

\bibitem{Zrenner1992}
A. Zrenner, P. Leeb, J. Sch{\"a}fler, G. B{\"o}hm, G. Weimann, J.M. Worlock, L.T. Florez, J.P. Harbison, Indirect excitons in coupled quantum well structures, {\it Surf. Sci.} {\bf 263}, 496-501 (1992).

\bibitem{Grosso2009}
G. Grosso, J. Graves, A.T. Hammack, A.A. High, L.V. Butov, M. Hanson, A.C. Gossard, Excitonic switches operating at around 100 K, {\it Nat. Photonics} {\bf 3}, 577-580 (2009).

\bibitem{Butov1995}
L.V. Butov, A. Zrenner, G. Abstreiter, A.V. Petinova, K. Eberl, Direct and indirect magnetoexcitons in symmetric In$_x$Ga$_{1-x}$As/GaAs coupled quantum wells, {\it Phys. Rev. B} {\bf 52}, 12153-12157 (1995).

\bibitem{Lefebvre2004}
P. Lefebvre, S. Kalliakos, T. Bretagnon, P. Valvin, T. Taliercio, B. Gil, N. Grandjean, J. Massies, Observation and modeling of the time-dependent descreening of internal electric field in a wurtzite GaN/Al$_{0.15}$Ga$_{0.85}$N quantum well after high photoexcitation, {\it Phys. Rev. B} {\bf 69}, 035307 (2004).

\bibitem{Fedichkin2015}
F. Fedichkin, P. Andreakou, B. Jouault, M. Vladimirova, T. Guillet, C. Brimont, P. Valvin, T. Bretagnon, A. Dussaigne, N. Grandjean, P. Lefebvre, Transport of dipolar excitons in (Al,Ga)N/GaN quantum wells, {\it Phys. Rev. B} {\bf 91}, 205424 (2015).

\bibitem{Fedichkin2016}
F. Fedichkin, T. Guillet, P. Valvin, B. Jouault, C. Brimont, T. Bretagnon, L. Lahourcade, N. Grandjean, P. Lefebvre, M. Vladimirova, Room-Temperature Transport of Indirect Excitons in (Al,Ga)N/GaN Quantum Wells, {\it Phys. Rev. Appl.} {\bf 6}, 014011 (2016).

\bibitem{Morhain2005}
C. Morhain, T. Bretagnon, P. Lefebvre, X. Tang, P. Valvin, T. Guillet, B. Gil, T. Taliercio, M. Teisseire-Doninelli, B. Vinter, C. Deparis, Internal electric field in wurtzite ZnO/Zn$_{0.78}$Mg$_{0.22}$O quantum wells, {\it Phys. Rev. B} {\bf 72}, 241305(R) (2005).

\bibitem{Kuznetsova2015}
Y.Y. Kuznetsova, F. Fedichkin, P. Andreakou, E.V. Calman, L.V. Butov, P. Lefebvre, T. Bretagnon, T. Guillet, M. Vladimirova, C. Morhain, J.-M. Chauveau, Transport of indirect excitons in ZnO quantum wells, {\it Opt. Lett.} {\bf 40}, 3667-3670 (2015).

\bibitem{Geim2013}
A.K. Geim, I.V. Grigorieva, Van der Waals heterostructures, {\it Nature} {\bf 499}, 419-425 (2013).

\bibitem{Ye2014}
Ziliang Ye, Ting Cao, Kevin O’Brien, Hanyu Zhu, Xiaobo Yin, Yuan Wang, Steven G. Louie, Xiang Zhang, Probing excitonic dark states in single-layer tungsten disulphide, {\it Nature} {\bf 513}, 214-218 (2014).

\bibitem{Chernikov2014}
Alexey Chernikov, Timothy C. Berkelbach, Heather M. Hill, Albert Rigosi, Yilei Li, Ozgur Burak Aslan, David R. Reichman, Mark S. Hybertsen, Tony F. Heinz, Exciton Binding Energy and Nonhydrogenic Rydberg Series in Monolayer WS$_2$, {\it Phys. Rev. Lett.} {\bf 113}, 076802 (2014).

\bibitem{Calman2016}
E.V. Calman, C.J. Dorow, M.M. Fogler, L.V. Butov, S. Hu, A. Mishchenko, A.K. Geim, Control of excitons in multi-layer van der Waals heterostructures, {\it Appl. Phys. Lett.} {\bf 108}, 101901 (2016).

\bibitem{Rivera2015}
% MoSe2-WSe2
Pasqual Rivera, John R. Schaibley, Aaron M. Jones, Jason S. Ross, Sanfeng Wu, Grant Aivazian,
Philip Klement, Kyle Seyler, Genevieve Clark, Nirmal J. Ghimire, Jiaqiang Yan, D.G. Mandrus, Wang Yao, Xiaodong Xu, Observation of long-lived interlayer excitons in monolayer MoSe$_2$-WSe$_2$ heterostructures, {\it Nat. Commun.} {\bf 6}, 6242 (2015).

\bibitem{Rivera2016}
% MoSe2-WSe2
Pasqual Rivera, Kyle L. Seyler, Hongyi Yu, John R. Schaibley, Jiaqiang Yan, David G. Mandrus, Wang Yao, Xiaodong Xu, Valley-polarized exciton dynamics in a 2D semiconductor heterostructure, {\it Science} {\bf 351}, 688-691, (2016). 

\bibitem{Nagler2017}
% MoSe2-WSe2
Philipp Nagler, Gerd Plechinger, Mariana V. Ballottin, Anatolie Mitioglu, Sebastian Meier, Nicola Paradiso, Christoph Strunk, Alexey Chernikov, Peter C.M. Christianen, Christian Sch{\"u}ller, Tobias Korn, Interlayer exciton dynamics in a dichalcogenide monolayer heterostructure, {\it 2D Mater.} {\bf 4}, 025112 (2017).

\bibitem{Fang2014}
% MoS2 - WSe2
Hui Fang, Corsin Battaglia, Carlo Carraro, Slavomir Nemsak, Burak Ozdol, Jeong Seuk Kang, Hans A. Bechtel, Sujay B. Desai, Florian Kronast, Ahmet A. Unal, Giuseppina Conti, Catherine Conlon, Gunnar K. Palsson, Michael C. Marting, Andrew M. Minor, Charles S. Fadley, Eli Yablonovitch, Roya Maboudian, Ali Javey, Strong interlayer coupling in van der Waals heterostructures built from single-layer chalcogenides, {\it PNAS} {\bf 111}, 6198 (2014).

\bibitem{Latini2017}
%MoS2 - WSe2 with hBN
Simone Latini, Kirsten T. Winther, Thomas Olsen, Kristian S. Thygesen, Interlayer Excitons and Band Alignment in MoS$_2$/hBN/WSe$_2$ van der Waals Heterostructures, {\it Nano Lett.} {\bf 17}, 938-945 (2017).

\bibitem{Hong2014}
%MoS2/WS2
Xiaoping Hong, Jonghwan Kim, Su-Fei Shi, Yu Zhang, Chenhao Jin, Yinghui Sun, Sefaattin Tongay, Junqiao Wu, Yanfeng Zhang, Feng Wang, Ultrafast charge transfer in atomically thin MoS$_2$/WS$_2$ heterostructures, {\it Nat. Nano} {\bf 9}, 682-686 (2014).

\bibitem{Palummo2015}
%MoS2/WS2 and MoSe2/WSe2
Maurizia Palummo, Marco Bernardi, Jeffrey C. Grossman, Exciton Radiative Lifetimes in Two-Dimensional Transition Metal Dichalcogenides, {\it Nano Lett.} {\bf 15}, 2794-2800 (2015).

\bibitem{Bellus2015}
%MoSe2 and WS2
Matthew Z. Bellus, Frank Ceballos, Hsin-Ying Chiu, Hui Zhao, Tightly Bound Trions in Transition Metal Dichalcogenide Heterostructures, {ACS Nano} {\bf 9}, 6459-6464, (2015).

\bibitem{Mouri2017}
%MoSe2/MoS2
Shinichiro Mouri, Wenjing Zhang, Daichi Kozawa, Yuhei Miyauchi, Goki Edac, Kazunari Matsuda, Thermal dissociation of inter-layer excitons in MoS$_2$/MoSe$_2$ hetero-bilayers, {\it Nanoscale} {\bf 9}, 6674-6679 (2017).

\bibitem{Britnell2012}
L. Britnell, R.V. Gorbachev, R. Jalil, B.D. Belle, F. Schedin, A. Mishchenko, T. Georgiou, M.I. Katsnelson, L. Eaves, S.V. Morozov, N.M.R. Peres, J. Leist, A.K. Geim, K.S. Novoselov, L.A. Ponomarenko, Field-Effect Tunneling Transistor Based on Vertical
Graphene Heterostructures, {\it Science} {\bf 335}, 947-950 (2012). 

\end{references}
\end{document}